# The Effect of Aqueous Medium on Nucleobase Shape Resonances: Insights from Microsolvation

Jishnu Narayanan S J,[1] Divya Tripathi,[1] Idan Haritan,[2] and Achintya Kumar Dutta[1,*]

*1. Department of Chemistry, Indian Institute of Technology Bombay, Powai, Mumbai 400076*

*2. Alexander Kofkin Faculty of Engineering, Bar-Ilan University, Ramat Gan, 5290002 Israel*

**Abstract**

We have studied the effect of microhydration on the shape resonances of uracil nucleobase. The resonance parameters were determined using the resonance via Padé approach along with the efficient wave function-based EA-EOM-DLPNO-CCSD method. Our results showed that the uracil resonances become stabilized with an increase in the extent of microsolvation. The energy of the resonances decreased, and the lifetime increased as the number of water molecules surrounding uracil was increased. It showed that ten water molecules are sufficient to make the lowest shape resonance of uracil a bound radical anionic state. Our results also indicate that the lowest energy resonance state may become a bound state under bulk solvation.

*achintya@chem.iitb.ac.in

# 1. Introduction

Dissociative electron attachment (DEA) to DNA accounts for a significant amount of the damage to DNA when it is subjected to ionizing radiation.[1,2] The high-energy photons ionize the surrounding environment of the DNA, resulting in the formation of multiple secondary products, which include low-energy electrons (LEE), free radicals, etc.[3,4] The DEA to genetic materials is initiated by LEE attachment to the biomolecule through the formation of transient negative ions (TNIs).[3] These quasi-bound anions, which have a finite lifetime, are also called anion resonances. Once formed, the temporary-bound anions can decay via three major pathways: DEA, autodetachment, or relaxing to a bound anionic state (shown as process 2, process 3 and process 4, respectively, in Scheme 1). While autodetachment and the formation of bound radical anions do not result in DNA strand breaks, the DEA can cause significant DNA damage, including single-strand breaks (SSB), double-strand breaks (DSB), clustered damages, etc.[3–10] Autodetachment does not cause strand breaks, as the extra electron escapes to the continuum, returning the DNA to its original neutral state. Similarly, the formation of bound radicals does not lead to strand breaks, since bound radical anions of the genetic material do not have sufficient energy to cause strand breaks.[4] However, nucleobase-centered stable DNA radical anions may cause point mutations that are repairable by the base-excision repair mechanism.[5] Several experimental[11–18] and theoretical[19–31] studies have highlighted the role of anion resonances, localized on the nucleobase or the phosphate group of the DNA, in the DEA process. These findings suggest that the quasi-bound anion facilitates the transfer of the additional electron to the σ* orbital of the sugar-phosphate or sugar-nucleobase bonds, which leads to their rupture.[32] Strand breaks caused by DEA to DNA depend on the energy and lifetime of the resonance state formed. Therefore, characterizing the various TNIs involved in LEE-induced radiation damage is essential to elucidate the mechanism of the process.

Quasi-bound anions can be broadly classified into shape/one-particle (1p) and core-excited/two particle-one-hole (2p1h) resonances. The Shape resonances, as the name indicates, are formed when the incoming electron temporarily binds to the system due to the shape of the interaction potential barrier.[33] The additional electron occupies an unoccupied molecular orbital of the system. 2p1h resonances are formed when the incoming electron has sufficient energy to excite another electron from an occupied orbital and get attached to the resulting excited state of the neutral molecule. Hence, 2p1h resonances are generally high-lying (>4.0 eV) states as compared to shape resonances.[34] When one considers electron attachment-induced damage to DNA, both shape and core-excited resonances are of importance. Based on their electron-irradiation experiments on plasmid DNAs, Sanche and coworkers suggested that long-lived core-excited resonances play a major role in DEA to genetic materials.[3,15,32] However, several other studies have since shown that LEEs with energies <3 eV also cause damage to DNA.[35–37] This suggests that shape-resonances may also play a significant role in DEA to DNA since they are likely to initiate the process at such low energies.

TNIs have proved to be extremely difficult to characterize via experimental studies since they have lifetimes in the range of $10^{-15}-10^{-12}$ seconds.[38] The *in-silico* modeling of these states is also equally challenging. In one approach, solving the time-dependent Schrödinger equation is necessary to accurately model temporary bound anionic states.[33] However, such simulations on reasonably large systems are often practically unfeasible owing to the large timescales to which the wavepacket, representing the resonance state, needs to be propagated.[33] Alternatively, another approach involves employing the time-independent Schrödinger equation under the non-Hermitian quantum mechanics regime, using outgoing boundary conditions to study such states. In the non-Hermitian picture, resonances are associated with complex eigenvalues represented by Siegert energy expression: $E = E_R - \Gamma/2$.[33] Here, $E_R$ denotes the resonance energy (position), and $\Gamma$ is the resonance width, which is inversely proportional to the lifetime ($\tau = \hbar/\Gamma$) of the metastable state.[33] Techniques such as complex basis functions,[39–41] complex scaling approach, and complex absorbing potentials (CAP)[42–45] have been used to model TNIs of DNA model systems. The drawback associated with these methods is that they involve modification to the Hamiltonian of the system. This means that significant changes need to be made to the conventional electronic structure codes. An alternate approach involves analytically continuing the real eigenvalues to the complex plane to obtain energy and width. These methods are broadly known as stabilization approaches and do not require modification to standard quantum chemistry software packages.[34,46–51] Within these approaches, Moiseyev and coworkers introduced a novel method called resonances via Padé (RVP).[46–48,52] They have since shown that the RVP method successfully reproduced the resonance energy and lifetime of several species that included both atomic and molecular systems.[46,52,53] In this work, we have employed the RVP method to simulate the anion resonances of the genetic materials.

One should also note that theoretical simulations involving the entire DNA are practically not feasible, since DNA is a macromolecule. Model systems of genetic material could be used to overcome this hurdle. Nucleobases are the smallest model systems that are also relevant because LEE attachment to DNA occurs majorly via nucleobase-centered resonances.[13,54–56] Uracil, due to its smallest size among the nucleobases and structural similarity to the thymine nucleobase, makes it attractive for our calculations that involve computationally demanding wavefunction-based methods. Therefore, we have chosen uracil as the model system in this work. One can find several works, both experimental and theoretical, on the TNIs of gas-phase uracil in the literature.[34,44,53,57–62] Aflatooni et al. reported the electron attachment energies of the three lowest TNIs of uracil with the help of electron transmission spectroscopy (ETS).[58] Their results showed the resonance energies of $1\pi^*$, $2\pi^*$, and $3\pi^*$ states as 0.22, 1.58, and 3.83 eV, respectively. Several attempts have since been made to theoretically simulate the metastable anions of uracil.[34,44,53,61,62] The R-matrix calculations initially reported by Lucchese and coworkers showed significant deviations from the experimental values.[57] Further attempts using stabilization method at the Koopmans' level,[61] scattering calculations done using Schwinger multichannel method with pseudopotentials (SMCPP)[62] and R-matrix method,[63] and complex absorbing potential (CAP) combined with symmetry-adapted cluster configuration interaction (SAC-CI)[44] have fared relatively better. The more recent work by Fennimore et al. follows the stabilization approach, which employs the generalized Padé approximation (GPA) combined with the equation-of-motion coupled-cluster singles and doubles method for electron-attached states (EA-EOM-CCSD).[34] Bouskila et al. have also studied the shape resonances of uracil using the stabilization method.[53] Although they have also calculated the stabilization plot of uracil at EA-EOM-CCSD level, they have followed the resonance via Padé (RVP) method to analytically dilate the real eigenvalues to the complex plane.

DNA exists in a highly concentrated aqueous solution constituted by water, salts, biomolecules, cell organelles, etc., which is the cellular medium. DNA is also closely associated with histone proteins when it is packed inside the nucleus. One should expect the species in its surrounding medium to significantly influence electron attachment to DNA. Hence, it is essential to include the effect of the cellular environment in experiments as well as in theoretical simulations while probing DEA to genetic materials. However, the complexity of the surrounding medium of DNA makes it a challenging task. To include the effect of the aqueous environment, one can resort to a bottom-to-top approach where the complexity of the system is gradually increased from gas-phase uracil to microsolvated uracil-water complexes. Simulations on these systems could be considered along with results from bulk-solvated DNA model systems to obtain a comprehensive picture. One could find several works in the literature where DEA to microsolvated nucleobases are analyzed experimentally.[64,65] Mass spectrometric analysis of microhydrated uracil and thymine anions revealed that the hydrogen-bonded water molecules enhanced the nucleobases' resistance to DEA.[64] Kočišek and coworkers proposed that the additional kinetic energy acquired by the pyrimidine nucleobase is efficiently transferred to the surrounding solvent molecules, which prevents hydrogen loss from the N-H bond.[64] However, these studies using mass spectrometry do not shed light on the influence of water on the resonance parameters of the system.[64,66,67] Verlet and coworkers,[65] in their recent work, studied the effect of microsolvation on the $2\pi^*$ and $3\pi^*$ resonances of microhydrated uracil anions. Their two-dimensional photoelectron spectroscopic results showed that the quasi-bound anions of pyrimidine nucleobase get stabilized with solvation. In this work, we have considered uracil solvated with one water molecule (uracil($H_2O$)), two water molecules (uracil($H_2O$)$_2$), and ten water molecules (uracil($H_2O$)$_{10}$) to assess the influence of the aqueous environment on the resonance states.

## 2. Theory and Computational Details

The EA-EOM-CCSD method is an excellent tool to generate the electron-attached energy spectrum of small molecules.[68] However, due to its $O(N^6)$ scaling and high storage requirement, it is not practically feasible to use it for systems larger than gas-phase pyrimidine nucleobases. One can employ EA-EOM-CCSD under the DLPNO formalism (EA-EOM-DLPNO-CCSD), which has proved to be an efficient candidate for modeling electron attachment to larger systems.[69,70] In this work, we have used the EA-EOM-DLPNO-CCSD method to obtain the energy spectrum of uracil and its microsolvated analogs. The stabilization plots were generated using the EA-EOM-DLPNO-CCSD method with aug-cc-pVDZ+1s1p1d basis set. The additional 1s, 1p, and 1d diffuse functions were added only on the heavy atoms. The exponent for each of the newly introduced Gaussians was obtained by dividing the exponent of the corresponding most diffuse function existing in the aug-cc-pVDZ basis set by a factor of 2.0. The NORMALPNO truncation threshold was used for EA-EOM-DLPNO-CCSD.

The stabilization graph is the energy spectrum of the molecule plotted as the function of a parameter alpha ($\alpha$). It is constructed by dividing the exponents of the two most diffuse s, p, and d, (only s and p for hydrogen) functions by $\alpha$, which is varied from 0.2 – ~1.8. Compared to the original basis set, the scaled basis becomes contracted when $\alpha < 1.0$ and vice-versa when $\alpha > 1.0$, since the diffuseness of a Gaussian function is inversely proportional to the exponent. The stabilization plot for gas phase uracil is shown in Figure 1, with the stable regions highlighted. One can equate the small changes in the basis functions brought about by slowly varying $\alpha$ to introducing a small perturbation to the system. Since resonances behave as stationary states within the interaction region, the corresponding real eigenvalues are largely invariant with respect to $\alpha$.[33,47] The continuum eigenstates rapidly vary with $\alpha$, and the regions corresponding to these states in the stabilization plot are characterized by steep slopes.[33,47] Therefore, resonance states in the stabilization plot appear in the energy spectrum as stable regions. The stable region for a quasi-bound anionic state in the plot is also associated with two avoided crossings.[47] The data from the stable region is supplied to the RVP method to determine the $E_R$ and $\Gamma$. The output of the RVP method is then filtered using a clusterization technique to remove any unphysical results.[52] The reported resonance position and width correspond to the statistically most well-behaved branch for a particular resonance state. Further information regarding the RVP method can be found elsewhere.[46,48]

The geometry of gas-phase uracil was optimized using the RI-MP2 method and def2-TZVP basis set. Uracil microsolvated with water molecules can exist in several conformers due to the multiple hydrogen-bonding sites present. We have performed conformational sampling with the CREST[71] package to obtain suitable conformers for uracil($H_2O$) and uracil($H_2O$)$_2$. The energetically lowest four isomers of the former and ten isomers of the latter were taken from the conformers obtained. The equilibrium neutral geometries were then determined at the RI-MP2/def2-TZVP level of theory. Vertical electron affinity (VEA) of the selected conformers was calculated using the EA-EOM-DLPNO-CCSD method and aug-cc-pVDZ basis set. The isomer that formed the most stable anion after vertical electron attachment was chosen for further calculations. The geometry of uracil($H_2O$)$_{10}$ was obtained in a systematic manner as follows. Uracil was bulk-solvated with water molecules to generate a 40 Å box with VMD software.[72] Classical molecular dynamics simulations were performed as described in this work [73] to equilibrate the system. The final snapshot of the equilibrated trajectory was taken, and solvents other than the ten closest to the solute were removed. The system was then optimized at the RI-MP2/def2-TZVP level of theory, where the coordinates of water molecules were kept frozen. All quantum mechanical (QM) calculations were carried out with ORCA 5.0.3.[74,75]

## 3. Results and Discussions

### 3.1 Uracil in Gas-Phase

Before delving into the impact water molecules have on the nucleobase resonances, it is essential to assess the level of accuracy of the method used. Therefore, we have compared our results for the gas-phase uracil with previous works, especially those using GPA and RVP-EA-EOM-CCSD since both works are based on the stabilization approach. Our results for the energy and widths of uracil resonances are overestimated compared to those obtained at GPA and RVP-EA-EOM-CCSD methods (Table 1), except for the $2\pi^*$ resonance which is underestimated compared to that in GPA. The extent of overestimation varied from ~0.07 eV for $1\pi^*$ resonance to ~0.25 eV for $3\pi^*$ resonance, when compared with the results of Fennimore et al.[34] Our results also show a small deviation from the RVP-EA-EOM-CCSD results by Bouskila et al.[53] However, one should note that stabilization methods are extremely sensitive to the basis set used. The RVP-EA-EOM-CCSD results presented in Table 1 employ the cc-pVTZ+2s2p2d (aug-cc-pVTZ for hydrogen) basis set, whereas ours and GPA[34] are performed with the aug-cc-pVDZ+1s1p1d basis set. Although Bouskila et al. have proposed cc-pVTZ+2s2p2d as the best basis set for modeling the uracil resonances, they have also performed the calculations using the aug-cc-pVDZ+1s1p1d basis set.[53] At this level of theory, they obtained 0.738, 2.44, and 5.277 eV as the resonance energy for $1\pi^*$, $2\pi^*$, and $3\pi^*$-type resonances, respectively. The $1\pi^*$ and $3\pi^*$ resonance positions are similar to our value, whereas the $2\pi^*$ is underestimated by ~0.20 eV.

This difference could arise from two factors: the number of basis functions scaled to generate the stabilization plot and the DLPNO approximation (at NORMALPNO settings) in EA-EOM-DLPNO-CCSD. We have scaled the two most diffuse basis functions with s, p, and d angular momentum for the heavy atoms, whereas Bouskila et al..[53] scaled only the additional diffuse functions added to the basis set. The effect of DLPNO approximation on the resonance parameters was analyzed by Dutta and coworkers for cytosine.[76] They compared the position and width of the pyrimidine nucleobase at the three DLPNO truncation settings with the canonical RVP-EA-EOM-CCSD value with aug-cc-pVDZ+1s1p1d basis set. The TIGHTPNO,

NORMALPNO, and LOOSEPNO parameters define the truncation thresholds of DLPNO in decreasing order of accuracy and computational cost. These thresholds are TcutPairs, TcutPNO, TcutDo, and TcutMKN. Since the accuracy of the energy spectrum depends on these thresholds, the quality of the resonance energy and lifetime calculated via RVP is also influenced by these parameters. Readers are encouraged to refer to reference[77] for more details regarding DLPNO approximation. The results for cytosine indicate that resonance energies determined at the NORMALPNO setting were in reasonable agreement with canonical ones. Nevertheless, we expect our RVP-EA-EOM-DLPNO-CCSD/aug-cc-pVDZ+1s1p1d level of theory to overestimate the resonance positions by ~0.1 to ~0.4 eV, compared to Bouskila et al.'s work. When compared to experimental values obtained with electron transmission spectroscopy, the uracil $1\pi^*$ resonance energy in the present work is red-shifted by 0.46 eV. The $2\pi^*$ and $3\pi^*$ states are overestimated by larger magnitudes of 0.67 eV and 1.40 eV, respectively. As noticed in previous stabilization-based works,[34,76] the difference between computed and experimental values decreases when the differences between two resonances are considered.

The resonance lifetimes are generally more sensitive to the method and basis set than energies. To the best of our knowledge, experimental studies reporting uracil resonance lifetimes are absent in the literature. Hence, calculated resonance widths can only be compared with the previous theoretical results. The lifetime of a shape resonance state generally decreases as one moves from low-lying states to high-lying ones. This is because one-particle resonances become unstable with an increase in energy. Since the resonance width is inversely proportional to the lifetime, one should expect the width to increase as we move from $1\pi^*$ to $3\pi^*$ resonance in each of the systems considered. The resonance widths obtained in the present study are lower than those determined by both GPA[34] and RVP-EA-EOM-CCSD methods.[53] However, the trend in the TNI widths is as expected since it increases from 0.007 eV for the $1\pi^*$ state to 0.493 eV for the $3\pi^*$ state of uracil and is consistent with the previous reports. [34,53]

**3.2 Effect of Microhydration**

Several attempts have been made to theoretically model the effect of microsolvation on DEA to nucleobases. Almost all works[79–81] in the literature, except the previous one from our group[76] and Matsika and co-workers[82,83] employ electron scattering calculations. Smyth et al. showed that DEA cross-sections are enhanced when uracil exists in a 5-water cluster.[79] Their study focused on N-H σ* resonances of uracil, rather than the π* resonances studied in this work. Sieradzka and Gorfinkiel analyzed the effect of water molecules on the π* resonances of thymine.[80] They found that water molecules have a stabilizing or destabilizing effect depending upon the nature of the hydrogen-bonding interaction with thymine. Cornetta et al. considered the uracil-6water complex in their SMCPP level calculations to find that both 1π* and 2π* shape resonances became stabilized, compared to gas-phase uracil.[81] Although R-matrix- and SMCPP-based scattering calculations have offered detailed insights into the DEA process in small molecules, they suffer from limitations, such as inadequate inclusion of correlation effects. In stabilization-based methods, this could be overcome with the help of choosing an appropriate electronic structure method for generating the stabilization plot.

Table 2 presents the resonance positions and widths of microhydrated uracil. The resonance energy and lifetime are highly dependent on the geometry of the system.[82] When the monohydrated uracil molecule is considered, one could construct several conformers based on the nature of hydrogen bonding between the single water molecule and uracil. Matsika and coworkers determined the 1π* resonance parameters of four isomers of uracil($H_2O$) with the GPA method.[82] We also identified four distinct conformers of the complex through conformational sampling using CREST.[71] The energy and lifetime of the uracil TNIs should be averaged over the four conformers. However, as the number of water molecules increases, the number of possible conformers also increases. We obtained ten unique conformers for uracil($H_2O$)$_2$ system. Generating the stabilization plot for a system necessitates one to perform hundreds of single-point calculations at the EA-EOM-DLPNO-CCSD level. Hence, it is not practically possible to ensemble average the energy and width of the resonance states of microsolvated analogs of uracil over all distinct conformers. Therefore, we have chosen the conformer with the highest VEA for uracil-water complexes for the RVP calculations. In the CREST-generated conformers for uracil($H_2O$)$_{10}$, the water molecules formed a separate cluster that largely excluded uracil (Figure S5). One should expect the solvent molecules to interact with the nucleobase and form a solvation shell around it. Therefore, we had to obtain the geometry from the snapshot of an equilibrated classical molecular dynamics trajectory. Uracil($H_2O$)$_{10}$ is the largest microsolvated system that has been experimentally studied.[65] We were curious to determine if 10 water molecules are sufficient to make the lowest shape resonance of uracil a bound anionic state. Hence, we have chosen uracil($H_2O$)$_{10}$ as the largest system considered in this work. The optimized neutral geometries of uracil(H$_2$O), uracil(H$_2$O)$_2$, and uracil(H$_2$O)$_{10}$ are shown in Figure 2.

Table 2 shows that the addition of a single water molecule has a stabilizing effect on the three resonance states considered. The magnitude of stabilization varies from 1π* to 3π* states, with the largest redshift observed for the 3π* state (0.139 eV). When compared with gas-phase uracil, the resonance position of the 1π* state decreased by 0.053 eV, and the 2π* state got redshifted only by 0.032 eV. The monohydrated uracil isomer in our study is identical to isomer 3 from Matsika and coworkers' work.[82] They obtained a resonance energy of 0.60 eV, which is lower than our value (0.62 eV), and their decay width (0.104 eV) is larger than the width 0.003 eV obtained in this work using RVP method. This trend is similar to the observations for gas-phase uracil when our results were compared with GPA results.[82] The lifetime of the metastable anionic states also increased upon the addition of the water molecule. The 1π* state's lifetime increased from 101 fs in bare uracil to 195 fs in monohydrated uracil. Similar observations could be made for the remaining two states. The second resonance became long-lived by ~10 fs, whereas the lifetime of the third state increased only by ~1.0 fs, from uracil to uracil(H$_2$O). As can be seen in Figure 2., the water molecule in uracil(H$_2$O) acts as both a hydrogen donor and acceptor. In such a scenario, a previous scattering study by Sieradzka and Gorfinkiel on microsolvated thymine showed that the 1π* state gets destabilized, whereas the 2π* resonance gets stabilized.[80] However, we found that the presence of water causes the three shape resonances of uracil to become stable. Since Matsika and coworkers have shown that the resonance positions of microsolvated nucleobases are also sensitive to the local geometry of the uracil-water complex[82,83], it may be possible that another isomer of uracil(H$_2$O) exhibit similar trends in resonance energies reported by Sieradzka and Gorfinkiel.[80] The natural orbitals corresponding to the quasi-bound anionic states of uracil in gas-phase and uracil(H$_2$O) are represented in Figure 3, which shows negligible changes in the qualitative nature of the resonances upon microhydraton.

The addition of the second water molecule to monohydrated nucleobase resulted in stabilization of greater magnitude (see Table 2) when compared to the extent of stabilization between uracil and uracil(H$_2$O) resonances. Among the three resonance states of uracil(H$_2$O)$_2$, the 3π* state has the largest redshifts in width and energy. Compared to monohydrated uracil, the energy and width of the highest shape resonance of uracil(H$_2$O)$_2$ got stabilized by 0.31 eV and 0.045 eV, respectively. The extent of resonance stabilization when compared with monohydrated uracil is the lowest for $1\pi^*$ state of uracil$(H_2O)_2$, with a redshift of 0.05 eV for energy and 0.001 eV for the width of the resonance. However, due to the inverse relation of resonance width to its lifetime, uracil$(H_2O)_2$ $1\pi^*$ state's lifetime is 140 fs larger than that of monohydrated uracil. The natural orbitals corresponding to uracil$(H_2O)_2$ resonances are shown in Figure 4.

Solvents surrounding uracil in the bulk medium would offer a cage effect, which may offer additional stabilization to nucleobase TNIs. The cage formed by water molecules could also aid in the effective dissipation of additional energy from the nucleobase, as Kočišek and coworkers have suggested.[64] As shown in Table 2, the addition of the first and second water molecules successively stabilizes the three shape resonances of uracil. Microhydration of the pyrimidine nucleobase may provide a glimpse into the impact of the bulk environment on TNIs. Nevertheless, uracil$(H_2O)$ and uracil$(H_2O)_2$ are inadequate to fully capture the cage effect of water molecules in the first solvation shell of uracil. One can resolve this issue by solvating the nucleobase with a sufficient number of solvents. Our RVP-EA-EOM-DLPNO-CCSD results for the uracil$(H_2O)_{10}$ system demonstrate remarkable stabilization of the quasi-bound anions in aqueous medium. We obtained a resonance energy of about 0.05 eV for the lowest resonance of uracil$(H_2O)_{10}$, which is a significant redshift of 0.53 eV compared to uracil$(H_2O)_2$. One should also note that we have performed the calculations using the aug-cc-pVDZ+1s1p1d basis set. Dutta and coworkers observed a redshift of 0.22 eV and 0.28 eV for the $1\pi^*$ resonance of cytosine and cytosine$(H_2O)$, respectively while comparing aug-cc-pVDZ+1s1p1d results to aug-cc-pVTZ+1s1p1d. This means that, in their case, increasing the basis set dimension makes the calculated value of $E_R$ closer to the experimental value. Therefore, we expect the $1\pi^*$ state of uracil to become bound in the presence of 10 water molecules if one determines the resonance energies with a larger basis set. To verify the same, we calculated the VEA of uracil$(H_2O)_{10}$ at EA-EOM-DLPNO-CCSD/aug-cc-pVTZ+1s1p1d level of theory. As expected, the ground state of uracil$(H_2O)_{10}$ anion had vertical detachment energy (VDE) of ~0.1 eV, which makes it a bound anionic state. In the aug-cc-pVDZ+1s1p1d basis set, the $2\pi^*$ resonance of uracil$(H_2O)_{10}$ also gets stabilized to a similar extent (0.52 eV) as the $1\pi^*$ state, compared to uracil in the presence of two water molecules. The energy of the third shape resonance (4.54 eV) also decreases compared to uracil$(H_2O)_2$ (4.78 eV), although the redshift is smaller than $1\pi^*$ and $2\pi^*$ state. This trend is also visible in the decay width for all three shape resonances (Table 2). The uracil($H_2O$)$_{10}$ $1\pi^*$ resonance has a lifetime larger by 270 fs compared to that of uracil($H_2O$)$_2$ in the aug-cc-pVDZ+1s1p1d basis set and becomes a bound state on going to the aug-cc-pVTZ+1s1p1d basis set. The $2\pi^*$ and $3\pi^*$ states of uracil($H_2O$)$_{10}$ have lifetimes of 42 fs and 4.2 fs, respectively, in the aug-cc-pVDZ+1s1p1d basis set.

The natural orbitals corresponding to the three shape resonances of uracil($H_2O$)$_{10}$ are represented in Figure 4. Upon careful examination, the natural orbitals corresponding to the $2\pi^*$ and $3\pi^*$ resonances indicate that the portion of the additional electron density also exists on the water molecules.

Verlet and coworkers showed that $2\pi^*$ and $3\pi^*$ resonances of uracil would become bound radical anion states under bulk solvation.[65] One should note that their experiments involved microsolvated uracil systems in their anion geometry. However, our work involves the metastable anions of uracil-water complexes formed by vertical electron attachment. Although direct comparison is not possible between the two results, the overall trend observed in resonance positions with respect to solvation remains the same. To theoretically determine the resonance energy and lifetime of bulk-solvated nucleobases accurately, one should consider the entire system at the QM level. However, this is not computationally feasible due to the large size of the system. A workaround is to perform the calculations on several microsolvated clusters and extrapolate to obtain the bulk values. However, calculating the resonance parameters using any of the available theoretical tools is not feasible for larger nucleobase-water clusters. Multi-level quantum mechanical/molecular mechanical (QM/MM) methods are often employed to tackle this issue. For uracil-water systems, the nucleobase could be treated at the QM level and the water molecules at the MM level. However, the QM region is prone to overpolarization when QM/MM is used with non-polarizable force fields and electrostatic embedding. One could perform QM/MM calculations with the EA-EOM-DLPNO-CCSD method at the QM level to generate the stabilization plot for uracil-water systems. However, the result would be identical to considering solvents as point charges and performing the calculation at the same level of theory. Hence, to assess the effect of the QM/MM model on the position and decay width of the metastable states, we considered the water molecules in uracil($H_2O$)$_{10}$ as point charges, whose values were taken from the TIP3P water model. The $1\pi^*$ state became a bound state with VEA of 0.36 eV as the point charges were introduced instead of water, in uracil($H_2O$)$_{10}$. Compared to the full QM results, $2\pi^*$ and $3\pi^*$ resonance positions redshifted by 0.37 and 0.94 eV, respectively. The two resonances also became long-lived compared to uracil$(H_2O)_{10}$ results with water molecules also in the QM region (Table 2).

## 4. Conclusions

Our calculations on microhydrated uracil show that the surrounding environment has a significant impact on LEE attachment to DNA. The successive addition of two water molecules to uracil resulted in remarkable stabilization of the three shape resonances. The energy, as well as the decay width of the resonances, decreased with the extent of solvation. Moreover, the presence of ten water molecules is sufficient to make the lowest resonance a bound radical anion. This shows that at bulk solvation, the lowest shape resonances of nucleobases may become bound anionic states and may not participate in DEA-induced bond breaking. Based on our results, one should be cautious while employing multilevel QM/MM-based techniques with non-polarizable force fields while modeling solvation effects on quasi-bound anions, since the water molecules that are present as point charges offer overstabilization to the QM region. Our results provide a glimpse into the extent of influence the surrounding medium has on the negative ion resonances of the genetic material. In addition to shape resonances, the high-lying and long-lived 2p1h resonances also play a vital role in DEA. Work is in progress to study the effect of solvation on the core-excited resonances of DNA model systems.

**Supplementary Material**

Cartesian coordinates of the uracil in gas phase and under microsolvation, and stabilization graphs of microsolvated uracil molecules are provided in the supplementary material.

**Acknowledgments**

The authors acknowledge the support from the IIT Bombay, CRG (Project no. CRG/2018/001549), and Matrix project of DST-SERB (Project No. MTR/2021/000420), CSIR-India (Project No. 01(3035)/21/EMR-II), UGC-India, DST-Inspire Faculty Fellowship (Project no. DST/INSPIRE/04/2017/001730), Prime Minister's Research Fellowship, ISRO (Project No. RD/0122-ISROC00-004) for financial support. IIT Bombay super computational facility, and C-DAC Supercomputing resources (Param Smriti, Param Brahma, Param Rudra) for computational time. AKD acknowledges the research fellowship funded by the EU

**Conflict of Interest:**

The authors have no conflicts to disclose.

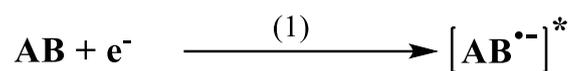

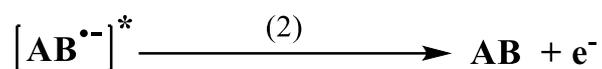

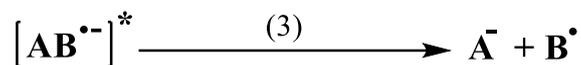

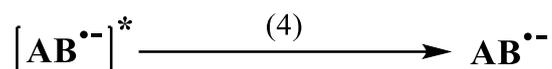

Scheme 1. Schematic representation of the three possible processes that can occur after the formation of a TNI. Here, AB is the molecule to which electron attachment occurs resulting into a metastable radical anion $[AB^{\bullet-}]^*$. In process 2, the extra electron gets autodetached to leave the neutral molecule. DEA occurs in process 3, where the TNI dissociates to generate $A^-$ and $B^\bullet$. The opposite process is also possible, where $B^-$ and $A^\bullet$ are formed upon dissociation. In process 4, the quasi-bound anion relaxes to the stable radical anion $AB^{\bullet-}$.

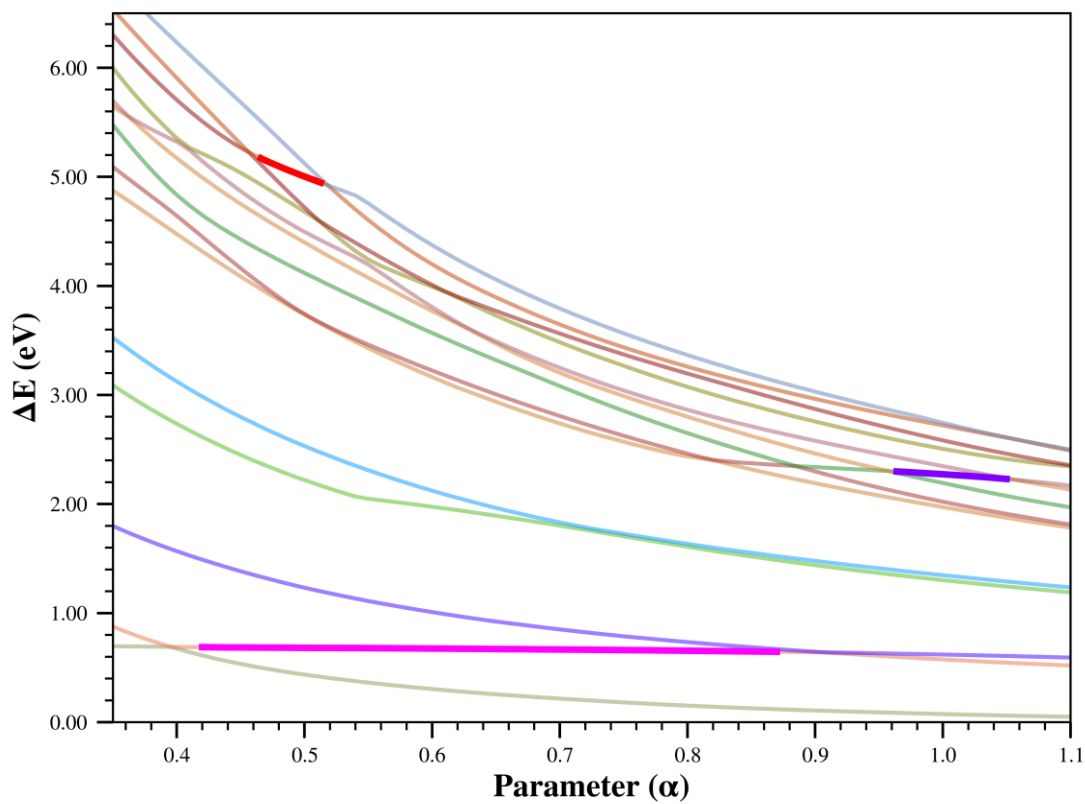

Figure 1. Stabilization plot for uracil anion in gas phase at EA-EOM-DLPNO-CCSD/aug-cc-pVDZ+1s1p1d level of theory. The stable regions corresponding to $1\pi^*$, $2\pi^*$, and $3\pi^*$ are separately highlighted in pink, violet, and red colors, respectively.

Table 1. Comparison of the energy and width of the resonance states of uracil calculated in the current work with previous theoretical results. All values are given in eV.

| Resonance state | 1π* | | 2π* | | 3π* | |
| --- | --- | --- | --- | --- | --- | --- |
| Method | $E_R$ | $\Gamma$ | $E_R$ | $\Gamma$ | $E_R$ | $\Gamma$ |
| RVP-EA-EOM-DLPNO-CCSD/aug-cc-pVDZ+1s1p1d [a] | 0.68 | 0.007 | 2.25 | 0.059 | 5.23 | 0.493 |
| RVP-EA-EOM-CCSD/aug-cc-pVDZ+1s1p1d[53] | 0.738 | 0.014 | 2.440 | 0.170 | 5.277 | 0.673 |
| RVP-EA-EOM-CCSD/cc-pVTZ+2s2p2d[53,b] | 0.597 | 0.014 | 2.183 | 0.14 | 4.858 | 0.657 |
| GPA-EA-EOM-CCSD[34,78] | 0.61 | 0.02 | 2.28 | 0.07 | 4.98 | 0.34 |
| CAP/SAC–CI[44] | 0.57 | 0.05 | 2.21 | 0.10 | 4.82 | 0.58 |
| Stab-DFT[61] | 0.36 | 0.05 | 1.75 | 0.10 | 4.52 | 0.23 |
| SMCPP[62] | 0.14 | 0.005 | 1.76 | 0.15 | 4.83 | 0.78 |
| R matrix (2009)[63] | 0.13 | 0.003 | 1.94 | 0.17 | 4.95 | 0.38 |
| R matrix (2004)[57] | 2.27 | 0.21 | 3.51 | 0.38 | 6.50 | 1.03 |
| Experiment[58] | 0.22 | | 1.58 | | 3.83 | |

[a] Current work

[b] aug-cc-pVTZ for the hydrogens

Table 2. The effect of microsolvation on the shape resonances of uracil determined using the RVP-EA-EOM-DLPNO-CCSD method. All values are given eV.

| Resonance state | $1\pi^*$ | | $2\pi^*$ | | $3\pi^*$ | |
|---|---|---|---|---|---|---|
| Molecule | $E_R$ | $\Gamma$ | $E_R$ | $\Gamma$ | $E_R$ | $\Gamma$ |
| Uracil | 0.68 | 0.007 | 2.25 | 0.059 | 5.23 | 0.493 |
| Uracil(H$_2$O) | 0.62 | 0.003 | 2.22 | 0.032 | 5.09 | 0.257 |
| Uracil(H$_2$O)$_2$ | 0.58 | 0.002 | 2.00 | 0.020 | 4.78 | 0.212 |
| Uracil(H$_2$O)$_{10}$ | 0.05 | 0.001 | 1.49 | 0.016 | 4.54 | 0.156 |
| Uracil(H$_2$O)$_{10}$ (water as point charges) | - | - | 1.12 | 0.012 | 3.60 | 0.064 |

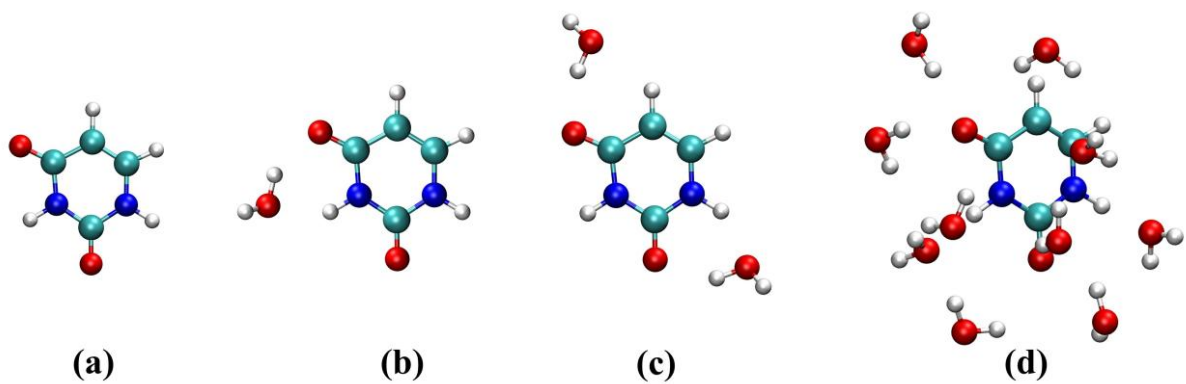

Figure 2. The optimized geometries of (a) gas-phase uracil, (b) uracil($H_2O$), (c) uracil($H_2O$)$_2$, and (d) uracil($H_2O$)$_{10}$ at RI-MP2/def2-TZVP level of theory.

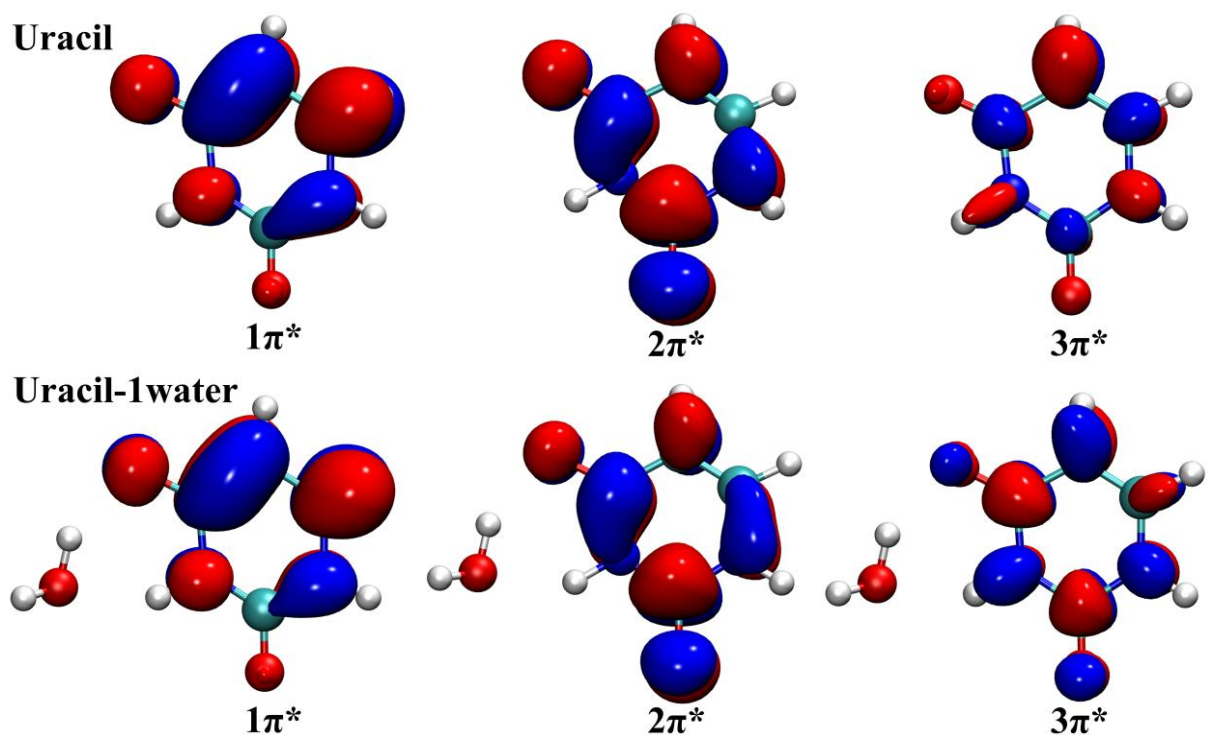

Figure 3. Natural orbitals corresponding to the three resonance states of uracil and uracil($H_2O$) at EA-EOM-DLPNO-CCSD/aug-cc-pVDZ+1s1p1d level of theory.

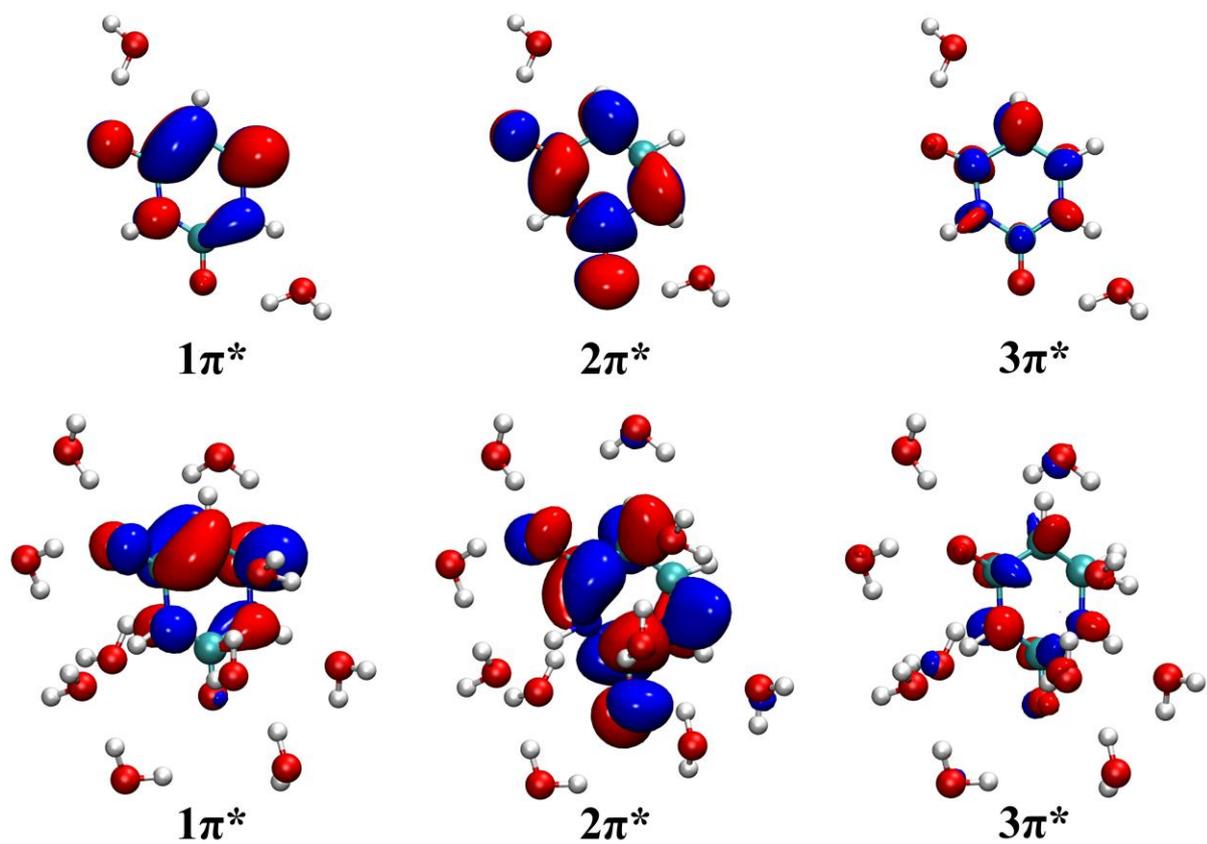

Figure 4. Natural orbitals corresponding to the three resonance states of uracil($H_2O$)$_2$ and uracil($H_2O$)$_{10}$ at the EA-EOM-DLPNO-CCSD/aug-cc-pVDZ+1s1p1d level of theory.